\documentstyle[twocolumn,prc,aps,floats,epsfig,graphics]{revtex}
\begin{document}
\draft

\twocolumn[\hsize\textwidth\columnwidth\hsize\csname
@twocolumnfalse\endcsname

\title{Systematic analysis of $\Delta I=4$ bifurcation in $A\sim$150
superdeformed nuclei: \\
 Active orbitals}
\author{I. M. Pavlichenkov \cite{IMP}, A. A. Shchurenkov}
\address{Russian Research Center "Kurchatov
Institute", Moscow, 123182, Russia}
\maketitle
\begin{abstract}
A simple criterion for the $\Delta I=4$ bifurcation is applied to
thirty superdeformed bands in the $A\sim$150 mass
region. The analysis allows to differentiate between active and
inactive for staggering single-particle states. The consideration is
based on additivity of the nonaxial hexadecapole moment,
which plays a key role in the phenomenon.
\end{abstract}
\vspace{0.5 cm}]

\narrowtext

The $\Delta I=4$ bifurcation, or the $\Delta I=2$ staggering, is a well
known mysterious phenomenon in the physics of  superdeformed (SD)
bands. It is observed as regular deviations of the $\gamma$--ray
energy differences $\Delta E_\gamma$ from the smooth behavior.
In spite of having the energy scale of tens of electron-volt the
phenomenon has been a topic of considerable interest and a lot of
the experimental and theoretical works have been devoted to it.
The reason lies in the unusual for the nuclear physics period of
oscillations, $\Delta I=4$, and their long and regular character.
Recently attention to this problem is called again by the
versatile analysis of the experimental data
undertaken in Refs.\ \cite{Lal,Has,Has1}.

Among other explanations \cite{HM,Mac,PF,Mag,MQ,Yan}
it has been suggested in
Ref.\cite{Pav} that the $\Delta I=4$ bifurcation is caused by the
coupling of rotation with single-particle motion in an axially
symmetric nuclei. The coupling is effected by the quadrupole and
hexadecapole two-body interactions through the term
\begin{equation}
V_{\rm {cpl}}= -\sum_{\lambda=2,4}\chi_{\lambda}\sum_{\mu\not=0}
Q_{\lambda\mu}{\cal Q}_{\lambda\mu},
\label{cpl}
\end{equation}
where $\chi_{\lambda}$ are the interaction strengths, $Q_{\lambda\mu}$
and ${\cal Q}_{\lambda\mu}$ are respectively the perturbative and
nonperturbative nonaxial multipole moments induced by rotation. The
separation of multipole moments in the two parts is explained by the two types
of the single-particle states involved in superdeformation:
low-$j$ natural and high-$j$ intruder orbitals. The former have
the natural parity and usually low angular momentum $j$.
They are weakly sensitive to the nuclear rotation and may be described
by the perturbation theory. Therefore, the perturbative parts are proportional
to nonaxial components of the total angular momentum $\bf I$:
\begin{equation}
{\cal Q}_{\lambda\pm 2}=\alpha_{\lambda 2}I^2_\pm, \hspace{5mm}
   {\cal Q}_{4\pm4}=\alpha_{44}I^4_\pm,
\label{mac}
\end{equation}
where $I_\pm=I_1\pm I_2$, $I_1$ and $I_2$ are the projections on the
coordinate axes perpendicular to the symmetry axis 3. On the other
hand, the intruder orbitals are very sensitive to rotation and, in addition,
each intruder state is affected differently by rotation. Therefore the
quantity $Q_{\lambda\mu}$  cannot be treated by the
perturbation theory.

The coupling (\ref{cpl}) distorts the rotational motion of an axially
symmetric nucleus. Accordingly, the effective rotational Hamiltonian
for an isolated band,
\begin{equation}
H_{\rm {eff}} = {\cal A}{\bf I}^2 + {\cal B}{\bf I}^4 + d(I^2_++I^2_-) +
    c(I^4_++I^4_-),
\label{Heff}
\end{equation}
contains nonadiabatic terms. The last two nonaxial terms split the
single band characteristic of an axial nucleus in a series of bands,
which correspond to the different orientation of the vector $\bf I$. The
Hamiltonian (\ref{Heff}) is adequate only for the description of the yrast
band, for which the angular momentum $\bf I$ is perpendicular to the
symmetry axis. All other bands are separated from the yrast one by a
large energetic gap caused by the small nonaxial deformation induced
by rotation. They are of no concern for us. The last term in (\ref{Heff})
proportional to $\chi_4$ is an essential ingredient
for the $\Delta I=2$ staggering. On the other hand, the term with the
operator $I^2_++I^2_-$ breaks a fourfold symmetry and makes the
staggering pattern irregular (see also Ref.\cite{Mag}). In other
words, the nonaxial terms of the Hamiltonian (\ref{Heff}) crimp
the rotational energy surface. The short wave crimps near the
stationary point (i.e., the axis of rotation) are important for staggering.
However the crimped surface does not yet solve the problem.
The staggering may exist if the stationary point is a minimum, which
happens for $c>0$. For the negative value of this parameter, the
staggering is absent in the yrast band, but exists in the uppermost
one because the transformation $c\rightarrow -c$ results in the
inversion of multiplet levels. The sign of $d$ is not important
for staggering.

\begin{figure*}[t]
\centerline{
\psfig{file=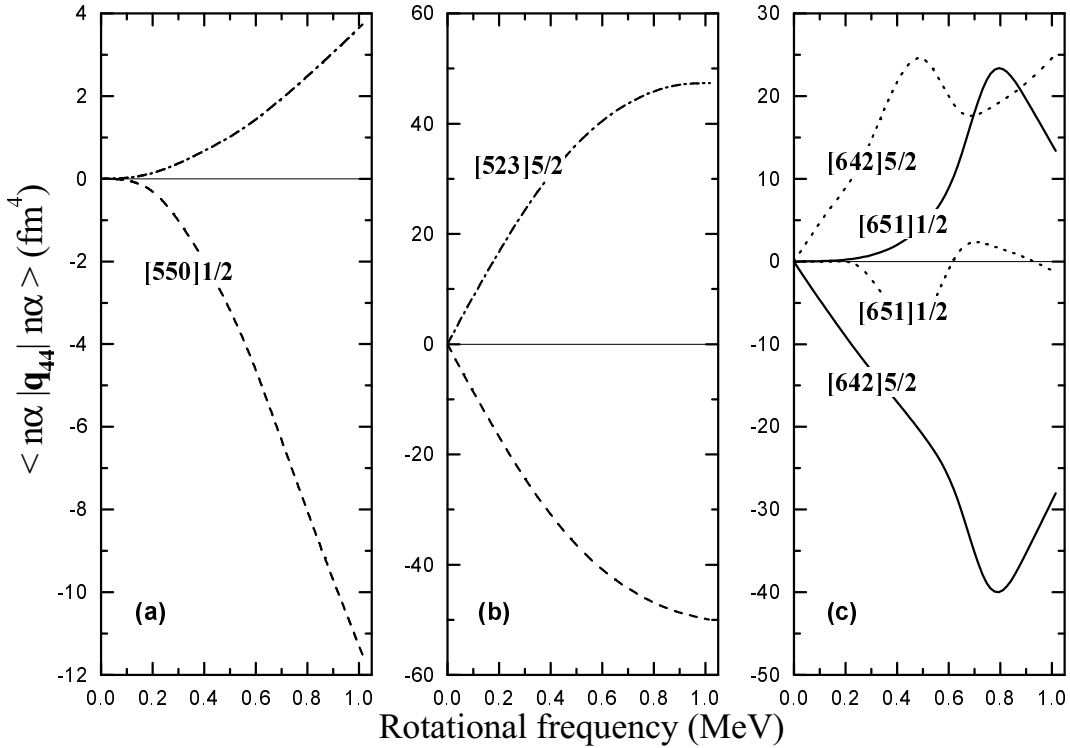,scale=0.5}}


\caption{Expectation values of the multipole moment $q_{44}({\bf r})$
plotted as a function of the rotational frequency $\omega$ for different
single-neutron orbitals. Deformation parameters ($\varepsilon_2=0.555,
\varepsilon_4=0.026$) correspond to the calculated minimum energy of
$^{149}$Gd(1) at $I\sim$ 40. The orbitals are labeled by
the asymptotic quantum numbers $n = [Nn_z\Lambda]\Omega$. Their
parity and signature $\alpha$
$(+,+\frac{1}{2})$, $(+,-\frac{1}{2})$, $(-,+\frac{1}{2})$, and
$(-,-\frac{1}{2})$ are indicated by solid, dotted, dashed and
dash-dotted lines, respectively. }
\label{qn}
\end{figure*}

The parameter $c$ involves the perturbative and nonperturbative
factors. For a nucleus with $Z$ protons ($\pi$) and $N$ neutrons
($\nu$) the necessary condition for the existence of the
$\Delta I=4$ bifurcation has the form
$$
c=-\chi_4\!\left[\Bigl(\frac{2Z}{A}\Bigr)^{2/3}\!\! Q_{44}(\pi)
      + \Bigl(\frac{2N}{A}\Bigr)^{2/3}\!\! Q_{44}(\nu)\right]
$$
\begin{equation}
\hspace{31pt}
\times\left[\Bigl(\frac{2Z}{A}\Bigr)^{2/3}\!\! \alpha_{44}(\pi) +
     \Bigl(\frac{2N}{A}\Bigr)^{2/3}\!\! \alpha_{44}(\nu)\right]>0,
\label{cond}
\end{equation}
where
\begin{equation}
Q_{44}(\tau)=
\sum_{n,\alpha}\langle n\alpha\tau|q_{44}|n\alpha\tau\rangle,
\hspace{20pt}\tau=\pi,\nu,
\label{addit}
\end{equation}
and the summation extends over all the occupied nonperturbative or,
{\it active} single-particle states having the quantum numbers $n$ and
the signature $\alpha$. We will use for $n$ the asymptotic quantum
numbers $[Nn_z\Lambda]\Omega$ of a nonrotating nucleus. The pairing effects
are neglected and the simplest shell model is used.
It should be noted that the condition (\ref{cond}) is
necessary but insufficient, due to a rather general form of the
rotation-single-particle interaction obtained from Eq. (\ref{cpl}).
More detailed information concerning this interaction is required to
get a sufficient condition and to reproduce staggering patterns.

The inequality (\ref{cond}) was used in Ref.\cite{Pav1} to check
the theory on eight SD bands in the $A\sim 150$ nuclei. The quantity
$\alpha_{44}$ was calculated in the anisotropic harmonic oscillator
(HO) potential and $Q_{44}$ was calculated in the limit of an
isolated intruder shell. It has been shown that the later quantity is the
fluctuating function of the number of the nucleons occupying intruder
orbitals. Thus the parameter $c$ may change sign and staggering
appears or disappears with the change of the intruder configuration
of a SD band. The correlation between the $\Delta I=4$ bifurcation
and the sign of $c$ has been found. However, the model in use is
not reliable for the SD bands because the intruder shells cannot be
treated as isolated.

In this paper, we report the results of the analysis of  thirty
SD bands, including those from Ref.\ \cite{Has}. We employ the
realistic modified oscillator (MO) potential by using the code GAMPN
\cite{Ragnar}. The $\kappa$ and $\mu$ parameters have been taken
from Ref.\cite{Haas}. The expectation values
$q_{44}(n\alpha\tau)=\langle n\alpha\tau|q_{44}|n\alpha\tau\rangle$
involved in $Q_{\lambda\mu}$ and ${\cal Q}_{\lambda\mu}$ are
calculated with the wave functions,
\begin{equation}
\psi_{n\alpha}=\sum_{lj\Omega}a^{n\alpha}_{lj\Omega}
      |N_{rot}lj\Omega\rangle,
\end{equation}
of the cranking potential, where $N_{rot}$ is the principal quantum
number in the stretched rotating basis. The small coupling between
different $N_{rot}$-shells is neglected.

\begin{figure*}[t]
\centerline{
\psfig{file=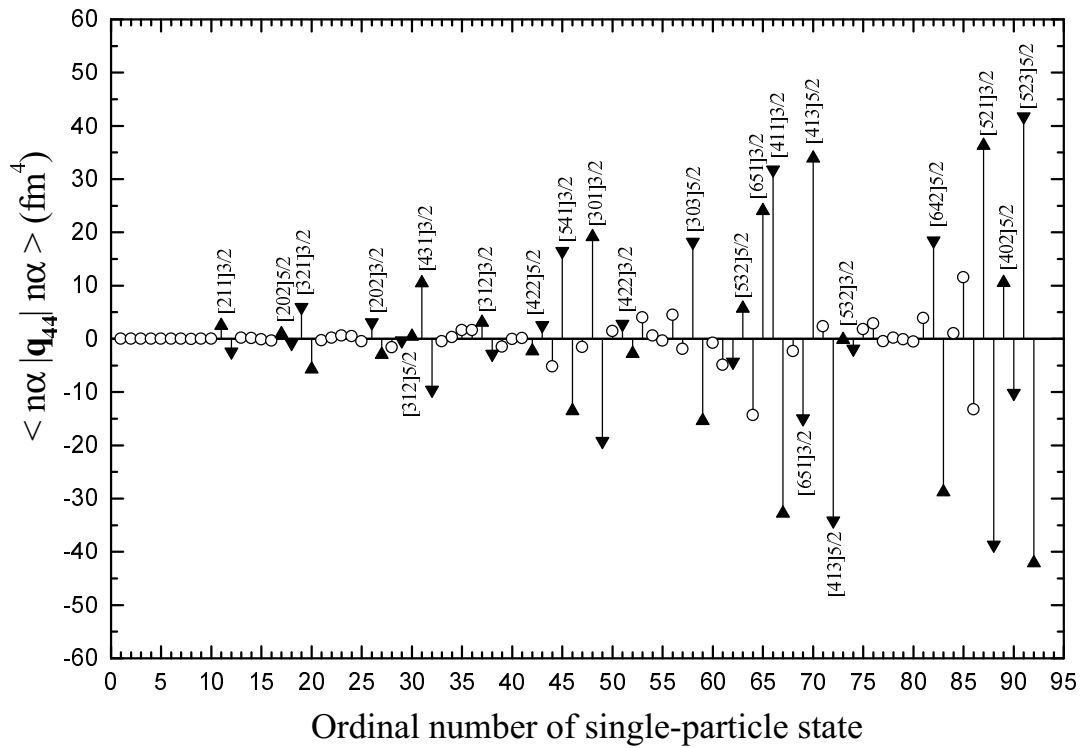,scale=0.5}}


\caption{Nonaxial hexadecapole moment of active (filled triangles)
and nonactive (empty circles) neutron orbitals as a function
of the ordinal number of a single-particle state in the cranking MO
potential for rotational frequency $\omega=0.8$ MeV. The asymptotic
quantum numbers $[Nn_z\Lambda]\Omega$ label the active states
with the signature $\alpha=+\frac{1}{2}$ (up triangles) and
$-\frac{1}{2}$ (down triangles). Deformation parameters are the
same as in Fig.\ref{qn}. }
\label{activ}
\end{figure*}

In order to single out the active orbitals we have investigated how 
the expectation values $q_{44}(n\alpha)$ depend on the rotational
frequency $\omega$. All the single-particle states occupied by
neutrons and protons in the $A\sim$150 nuclei have been 
considered. The following conclusions can be drawn.
({\it i}) There are the three different patterns of the
$\omega$-dependence, which are shown in Fig.\ref{qn}. Each of the
patterns is associated with the specific types of orbitals.
({\it ii}) The perturbative (Fig.\ref{qn}a) and nonperturbative
(Fig.\ref{qn}b) dependencies associate with inactive and active for
staggering orbitals respectively. These orbitals are the same for
neutrons and protons. ({\it iii}) The set of active orbitals is
supplemented with the inactive states interacting with the active ones.
The orbitals $\nu$[642]5/2 and $\nu$[651]1/2 placed above the
neutron gap at $N=80$ represent a classical example
of an avoid level crossing \cite{Hebb}. Figure \ref{qn}c shows
$q_{44}$ as a function of $\omega$ for the two signature branches of
these orbitals. According to the Strutinsky prescription, we use the
renormalization factor 1.27 to have the possibility of a comparison
with observed rotational frequencies. At low frequency
the [651]1/2 orbitals are inactive,
whereas at high frequency they involve large admixture of the active,
[642]5/2, orbitals and have the nonperturbative dependence of
$q_{44}$. These interacting orbitals with the positive signature
carry considerable hexadecapole moment. Thus the removal or 
addition of a neutron in these states may change the inequality (\ref{cond}).
The orbitals $\pi$[402]1/2 and $\nu$[301]1/2 with both signatures
are active practically for all frequencies because their avoid
crossings with $\pi$[422]3/2 and $\nu$[303]5/2 occur at low
frequencies. Some other active states induced by an avoid crossing
are presented in Table \ref{tab4}. Their contribution to the
quantities $Q_{44}(\pi)$ or $Q_{44}(\nu)$ is moderate. All the
considered pairs of interacting states belong to the same
$N_{rot}$-shell. The coupling between different $N_{rot}$-shells
generates the avoid crossings, which change the moments
$q_{44}$ only slightly due to a small interaction.

The set of active for staggering states is shown in Fig.\ref{activ}.
These are not necessary the intruder orbitals but all have the
asymptotic quantum numbers $\Omega$=3/2 and 5/2. Compared
to inactive orbitals, active ones carry the large values of $q_{44}$,
which are nevertheless smaller that the hexadecapole $q_{40}$
moment of the single-particle states around the SD core of
$^{152}$Dy \cite{Sat}. As a rule, the moments of the states with the
same asymptotic quantum numbers have close absolute values and
opposite signs for different signatures. They almost offset each other.
Thus, the contribution of the active orbitals to the total nonaxial
hexadecapole moment has the same order as that of inactive ones.
Consequently the equilibrium deformation $\varepsilon_{44}$ at a
high rotational frequency is small \cite{Rag,Naz}.

\begin{table*}[t]
\caption{Total moment $Q_{44}$ of the neutron and proton active
orbitals, the sign of the coefficient $c$, and the staggering
significance $Y$ for the SD bands with the known equilibrium
deformation for three rotational frequencies $\omega$ (MeV).
The subscripts + and $-$ denote the sign of the signature $\alpha$.
The standard notation is used for intruder orbitals.}

\begin{tabular}{lcr@{\hspace{32pt}}r@{\hspace{32pt}}r@{\hspace{32pt}}cc}
&Configuration&\multicolumn{3}{r@{\hspace{76pt}}}{$Q_{44}\ \ (\hbar/M\omega_0)^2$} &Sign &\\
Band&(relative to $^{149}$Gd(1))&\multicolumn{1}{r@{\hspace{27pt}}}{$\omega=0.4$} 
&\multicolumn{1}{r@{\hspace{27pt}}}{$\omega=0.6$}&\multicolumn{1}{r@{\hspace{27pt}}}{
$\omega=0.8$} &of $c$ &$Y$\\
\tableline
\vspace{-6pt}
      &        &       &     &      &     &   \\[1pt]
$^{147}$Gd(2) & $\nu[642\frac{5}{2}]_+^{-1}\nu[651\frac{1}{2}]_+^{-1}$ &
                           0.891 & 1.161 & 1.194 & $c>0$ &0.05 \\ [2pt]
$^{147}$Gd(3) & $\nu[651\frac{1}{2}]_+^{-1}\nu 7^{-1}_1$ &
                           0.389& 0.485 & 0.443&$c>0$& 0.15 \\ [2pt]
$^{147}$Gd(4) & $\nu[411\frac{1}{2}]_+^{-1}\nu[651\frac{1}{2}]_+^{-1}$ &
                     0.284 & 0.195 &-- 0.341 &$c<0$& 0.16 \\ [2pt]
$^{148}$Gd(1) &$\nu[651\frac{1}{2}]_+^{-1}$&
                     0.235 & 0.199 &-- 0.264 &$c<0$&0.23 \\ [2pt]
$^{148}$Gd(3) &$\nu 7^{-1}_1$&
                     0.197 & 0.361&0.270 &$c>0$& 0.10 \\ [2pt]
$^{148}$Gd(4) &$\nu[642\frac{5}{2}]_+^{-1}$&
                      1.011 & 1.131 & 2.059 &$c>0$& 0.98 \\ [2pt]
$^{148}$Gd(6) & $\nu[411\frac{1}{2}]_+^{-1}$ &
                          0.337 & 0.369 &0.683 &$c>0$& 3.1 \\ [2pt]
$^{149}$Gd(1) &                                &
                           0.321 & 0.227 & 0.505 &$c>0$ & 2.3 \\ [2pt]
$^{149}$Gd(5) &$\nu[402\frac{5}{2}]_+\nu[651\frac{1}{2}]_+^{-1} $&
                           0.344 & 0.456 & 0.600 &$c>0$ &   \\ [2pt]
$^{149}$Gd(6) &$\nu[402\frac{5}{2}]_-\nu[651\frac{1}{2}]_+^{-1}$&
                           --0.112 &-- 0.297 &-- 0.516 &$c<0$ &   \\ [2pt]
$^{150}$Gd(1) & $\nu 7_2$ &
                       --0.079 & 0.112 & 0.625 &$c>0$ & 0.50 \\ [2pt]
$^{150}$Gd(4a) & $\nu[402\frac{5}{2}]_+$ &
                            0.485 & 0.316 &0.864 &$c>0$ & \\ [2pt]
$^{150}$Gd(4b) &$\nu[402\frac{5}{2}]_-$   &
                             0.025 & --0.442 &-- 0.265 &$c<0$ & 1.0 \\ [2pt]
$^{151}$Tb(1) &$\pi 6_3\nu 7_2$&
                           0.776 & 1.320 & 1.477 &$c>0$ &   \\ [2pt]
$^{152}$Dy(1) &$\pi 6_3\pi 6_4\nu 7_2$&
                           --0.040 & 0.618 & 0.784 &$c>0$ &   \\ [2pt]
$^{153}$Dy(1) &$\pi 6_3\pi 6_4\nu 7_2\nu 7_3$&
                           0.915 & 1.317 & 1.382 &$c>0$ &   \\ [2pt]
\end{tabular}
\label{tab1}
\end{table*}

To explain these findings let us consider a cranking isolated
$j$-shell with the Hamiltonian
\begin{equation}
H_j=H_{j0} + \epsilon j^2_3 - \omega j_1,
\end{equation}
where $H_{j0}$ is a spherical part and $\epsilon$ is proportional to an
axial deformation. Assuming the small rotational frequency $\omega$,
we shall use perturbation theory with the unperturbed function in the
signature representation
\begin{equation}
u_{j\Omega\alpha}=(|j\Omega\rangle +
           e^{i\pi(j-\alpha)} |j-\Omega\rangle)/\sqrt{2}.
\end{equation}
It is easy to see that the only expectation values of $q_{44}(\bf r)$,
which are proportional to $\omega$, are those for the states with
$\Omega$=3/2 and 5/2:
\FL
\begin{equation}
q_{44}(j\ 3/2\ \alpha)\!=\!-q_{44}(j\ 5/2\ \alpha)\!=
   \!\frac{\omega}{\epsilon}\langle\|q_4\|\rangle f(j)e^{i\pi(j-\alpha)},
\label{isol}
\end{equation}
where the form of the positive function $f$ is inessential for us. For other
$\Omega$, the first non-vanishing contribution to
$q_{44}(j\Omega\alpha)$ is proportional to the higher powers of
$\omega$. The equality (\ref{isol}) explains the signature
dependence of the values $q_{44}$ for almost all active orbitals
shown in Fig.\ref{activ} because they are mostly high-j intruder or
high-j ones with a rather good quantum number $j$. An anomalous
signature dependence is observed for the five states with small $j$.

In a similar way as in Ref.\ \cite{Pav1}, we now check the criterion
(\ref{cond}) bearing in mind that the third multiplier is negative
for all the bands considered below. Really, straightforward
calculation within the MO potential show that the perturbative
quantity $\alpha_{44}$ is negative as in the case of the HO
model\ \cite{Pav1}. Thus we will be interested only in the sign of
the second multiplier
\begin{equation}
Q_{44}=\Bigl(\frac{2Z}{A}\Bigr)^{2/3}\! Q_{44}(\pi)
      + \Bigl(\frac{2N}{A}\Bigr)^{2/3}\! Q_{44}(\nu),
\label{qu}
\end{equation}
where the moments $Q_{44}(\pi)$ and $Q_{44}(\nu)$ are
calculated by using additivity of contributions from individual
orbitals according to Eq.\ (\ref{addit}).
The main difficulty in their calculation is the nuclear equilibrium
deformation, since the shape trajectories in the
$(\varepsilon,\varepsilon_4)$-plane are known for limited number
of SD bands. Starting with these bands, we give in Table\ \ref{tab1}
the estimated values of $Q_{44}$ for three rotational frequencies.
The corresponding parameters $\varepsilon$ and $\varepsilon_4$
have been taken from Refs.\ \cite{Thei} ($^{147}$Gd), \cite{Sav,Rag1}
($^{148}$Gd), \cite{Flib} ($^{149}$Gd), \cite{Nazar} ($^{150}$Gd,
$^{151}$Tb, $^{152,153}$Dy), and \cite{deFra}
($^{150}$Gd(4a,4b)). The present analysis
has an advantage because the sign of $c$
can be compared with the staggering significance $Y$
found in Ref.\ \cite{Has}. According to this work, the significance
is equal to the mean staggering amplitude divided by its
uncertainty. It is highly unlikely that all the bands with the
significance $Y>2$ exhibit the $\Delta I=4$
bifurcation only because of statistical fluctuations in the
$\gamma$-ray energy measurements. In particular, the independent
measurements of the $^{149}$Gd(1) staggering conclusively
demonstrate the existence of the effect. We use this band as a
reference for the single-particle structures of all the bands
studied in this work.

\begin{table*}[t]
\caption{The same as in Table\ \ref{tab1} for the SD bands, which
deformations $\varepsilon$ and $\varepsilon_4$ are estimated by using
the additivity principle. }

\begin{tabular}{lccccr@{\hspace{32pt}}cc}
         &Configuration&$\omega$&       &      &\multicolumn{1}{r@{\hspace{22pt}}}{$Q_{44}\hspace{12pt}$} &Sign &  \\
Band&(relative to $^{149}$Gd(1))&(MeV)&$\varepsilon$&$\varepsilon_4$ &
                             \multicolumn{1}{r@{\hspace{22pt}}}{$(\hbar/M\omega_0)^2$} &of $c$ &$Y$ \\
\tableline
\vspace{-6pt}
      &       &     &       &     &      &     &   \\[1pt]
$^{147}$Eu(1) & $\pi[301\frac{1}{2}]_-^{-1}\nu[651\frac{1}{2}]_+^{-1}$ &
                         0.8& 0.554 & 0.043 & --0.117 &$c<0$&0.95 \\ [2pt]
$^{147}$Eu(3) & $\pi[301\frac{1}{2}]^{-2}\pi 6_3\nu[651\frac{1}{2}]_-^{-1} $ &
                          0.8& 0.573 & 0.047 &1.191 &$c>0$ &0.58 \\ [2pt]
$^{148}$Eu(1) & $\pi[301\frac{1}{2}]_-^{-1}$ &
                          0.8& 0.564 & 0.042 &0.617 &$c>0$ & 2.3 \\ [2pt]
$^{148}$Gd(5) &$\pi[301\frac{1}{2}]^{-2}\pi 6_3\pi 6_4\nu[411\frac{1}{2}]^{-2}\nu 7_2$&
                          0.8& 0.619\tablenote{Deformation parameters 
are taken from \cite{Sav}}$\hspace{-4.3pt}$ & 0.037$^{\rm a}\hspace{-4.3pt}$ & 0.010 &$c>0$ & 2.8 \\ [2pt]
$^{150}$Gd(2) &$\pi[301\frac{1}{2}]^{-2}\pi 6_3\pi 6_4\nu 7_2$&
                          0.8& 0.608 & 0.042 & 0.743 &$c>0$ & 0.14 \\ [2pt]
$^{150}$Gd(8a) &$\pi([301\frac{1}{2}]_-^{-1}\pi 6_3\nu[402\frac{5}{2}]_-$&
                          0.8& 0.553 & 0.029 & 0.549 &$c>0$ & 0.13   \\ [2pt]
$^{150}$Gd(8b) &$\pi[301\frac{1}{2}]_-^{-1}\pi 6_3\nu[402\frac{5}{2}]_+$&
                           0.8& 0.553 & 0.029 &1.719 &$c>0$ &   \\ [2pt]
$^{151}$Gd(1a) &$\nu7_2\nu[402\frac{5}{2}]_+$&
                           0.6& 0.545 & 0.012 & 0.480 &$c>0$ & 1.8   \\ [2pt]
$^{151}$Gd(1b) &$\nu7_2\nu[402\frac{5}{2}]_-$&
                           0.6& 0.545 & 0.012 &--0.280 &$c<0$ & 0.25   \\ [2pt]
$^{152}$Tb(1) &$\pi 6_3\nu 7_2\nu[402\frac{5}{2}]_+$&
                           0.6& 0.554 & 0.012 & 1.051 &$c>0$ &   \\ [2pt]
$^{152}$Tb(2) &$\pi 6_3\nu 7_2\nu[402\frac{5}{2}]_-$&
                           0.6& 0.554 & 0.012 & 0.276 &$c>0$ &   \\ [2pt]
$^{152}$Dy(4) &$\pi 6_3\pi 6_4\nu[402\frac{5}{2}]_-$&
                           0.6& 0.551 & 0.015 & --0.387 &$c<0$ &   \\ [2pt]
$^{152}$Dy(5) &$\pi 6_3\pi 6_4\nu[402\frac{5}{2}]_+$&
                           0.6& 0.551 & 0.015 & 0.394 &$c>0$ &   \\ [2pt]
$^{153}$Dy(2) &$\pi 6_3\pi 6_4\nu 7_2\nu[402\frac{5}{2}]_-$&
                           0.6& 0.562 & 0.009 & --0.362 &$c<0$ &   \\ [2pt]
$^{153}$Dy(3) &$\pi 6_3\pi 6_4\nu 7_2\nu[402\frac{5}{2}]_+$&
                           0.6& 0.562 & 0.009 & 0.421 &$c>0$ &   \\ [2pt]
\end{tabular}
\label{tab2}
\end{table*}

Table\ \ref{tab2} presents the bands without calculated deformation.
In their analysis we have used the observation of Ref.\ \cite{Haas} that
the filling of any particular orbital always induced the same
deformation change in different nuclei. Subsequently
this feature has been explained by the additivity of quadrupole and
hexadecapole moments for SD bands in the $A\sim$150 mass
region \cite{Sat,Karl}. In a similar way as in the cited works we find
the deformation changes $\Delta\varepsilon$ and
$\Delta\varepsilon_4$, induced by a nucleon in the given state.
The corresponding values are presented in Table \ref{tab3} for the
two rotational frequencies. They are used to evaluate the
parameters $\varepsilon$ and $\varepsilon_4$ of the bands in
Table \ref{tab2}. The bands $^{150}$Gd(6a,6b) are not given in
the last table because the deformation changes induced by the
orbital $\nu[514]9/2$ are not known.

Tables \ref{tab1} and \ref{tab2} help to understand which property of
the single-particle structure is responsible for the $\Delta I=4$
bifurcation. First of all we would like to emphasize that the necessary
condition (\ref{cond}) is not violated in either of the
bands with the known staggering significance. This is not a trivial
fact because of the double cancellations in the expression $Q_{44}$:
the partial cancellation of the $q_{44}(n\alpha)$ values with
different signatures and the partial cancellation of the quantities
$Q_{44}(\pi)$ and $Q_{44}(\nu)$ for almost
all these bands. As a direct consequence of these cancellations,
the value $Q_{44}$ for some bands with the small significance $Y$
changes sign and turns negative for high rotational frequencies.
Besides the deformed shell model potential, the zero point of
$Q_{44}$ depends also on the frequency renormalization factor,
for which we take the conventional value 1.27. With such scaling
the criterion (\ref{cond}) seems to be not reliable for small
frequencies. Thus we use the high frequencies ($\omega=$ 0.6
and 0.8 MeV) to compare the staggering criterion with the
experimental significance.

While Tables \ref{tab1} and \ref{tab2} exhibit definitely the 
correlations between the sign of the parameter $c$
\begin{table}[h]
\caption{Deformation changes induced by an nucleon in the
fixed single-particle state. Only these orbitals are necessary
to estimate the deformation parameters of the SD bands
presented in Table \ref{tab2}. With the exception of the state
$\pi$[301]1/2, all the values have been extracted from pairs of bands
in adjacent nuclei with an extra nucleon in the indicated orbital. }

\begin{tabular}{crrrrc}
&\multicolumn{2}{c}{$\omega\!=\!0.6$MeV}&
\multicolumn{2}{c}{$\omega\!=\!0.8$MeV}\\
Orbital&$\delta\varepsilon$&$\delta\varepsilon_4$&
$\delta\varepsilon$&$\delta\varepsilon_4$&Source \\ [2pt]
\tableline
\vspace{-6pt}
      &        &       &        &         &   \\[1pt]
$\pi[301\frac{1}{2}]_+$&--0.013&--0.008&--0.014&--0.009&
                                          $^{149}$Gd(4) vs $^{150}$Gd(1)  \\ [1pt]
$\pi[301\frac{1}{2}]_-$&--0.012&--0.007&--0.013&--0.008&
                                           $^{149}$Gd(3) vs $^{149}$Gd(1) \\ [1pt]
$\pi[651\frac{3}{2}]_+$&0.009&--0.001&0.009&0.000&
                                           $^{151}$Tb(1) vs $^{150}$Gd(1)   \\ [1pt]
$\pi[651\frac{3}{2}]_-$&0.008&--0.003&0.009&--0.004&
                                           $^{152}$Dy(1) vs $^{151}$Tb(1)   \\ [1pt]
$\nu[651\frac{1}{2}]_-$&0.010&0.002&0.009&0.003&
                                      $^{149}$Gd(1) vs $^{148}$Gd(2) \\
                                        &0.011&--0.001&0.009&0.001&
                                      $^{148}$Gd(1) vs $^{147}$Gd(1)  \\ [1pt]
$\nu[770\frac{1}{2}]_+$&0.011&--0.006&0.012&--0.004&
                                      $^{150}$Gd(1) vs $^{149}$Gd(1)  \\ [1pt]
\end{tabular}
\label{tab3}
\end{table}
and the 
significance $Y$, they also show some discrepancies. 
The high positive value of $Q_{44}$ in the bands $^{147}$Gd(2) 
and $^{148}$Gd(4) is the consequence of the neutron 
hole in the state $\nu[642]5/2\ (\alpha=1/2)$, which
has according to Table\ \ref{tab4} the large negative $q_{44}$.
The same effect produces the orbital $\pi 6_3$ in the bands
$^{147}$Eu(3) and $^{150}$Gd(8a). The discrepancies observed
in the bands $^{147}$Gd(3) and $^{148}$Gd(3) are less evident.
Among the bands under study only these bands have the empty
state $\nu 7_1$. It is possible that the first intruder plays a
crucial role in the phenomenon (let us recall that the criterion
(\ref{cond}) is only necessary). This tentative conclusion is
confirmed by the nonstaggering bands  $^{150}$Gd(1,2),
$^{151}$Tb(1), and $^{152}$Dy(1), but it disagrees with the
staggering bands $^{148}$Gd(5) and $^{151}$Gd(1a). The first
intruder is blocked up by the second one, $\nu 7_2$, in these
bands (see also Ref.\ \cite{Flibo}). Let us note also that the first
proton intruder $\pi 6_1$ is blocked up in all the bands under study.

From a strictly logical point of view, better test of the inequality
(\ref{cond}) is provided by the pairs of the bands with
configurations that differ by a single nucleon occupying an
active or inactive orbitals. The filling of the
inactive orbital $\pi[301]1/2\ (\alpha=-1/2)$ does not change
$Q_{44}$. Therefore any pair of the identical bands
$^{147}$Eu(1)/$^{148}$Gd(1), $^{148}$Eu(1)/$^{149}$Gd(1)
has identical staggering properties. The same is true for the
identical bands $^{147}$Gd(4)/$^{148}$Gd(1) and $^{148}$Gd(6)/$^{149}$Gd(1), which
configurations are distinguished by a neutron in the state
$[411]1/2\ (\alpha=-1/2)$. Identical staggering properties
have the pair of bands $^{150}$Gd(2)/$^{152}$Dy(1)
differing in two protons in the state $[301]1/2$.
\begin{table*}[t]
\caption{ Expectation values $\langle\tau|q_{44}|\tau\rangle$
(in $(\hbar/M\omega_{0\tau})^2 $, $\tau=\pi,\nu$) of the active
states near the Fermi surface involved in the configurations of the bands with the known
staggering significance $Y$. All the quantities are calculated for
the fixed rotational frequency $\omega$=0.8\, MeV. A blank space
means that the corresponding orbital is inactive in the given band. }

\begin{tabular}{lrrrrrrrrrrr}
&\multicolumn{4}{c}{Proton states}&
\multicolumn{7}{c}{Neutron states}\\
Band&$\pi[301\frac{1}{2}]_+$&
           $\pi[541\frac{1}{2}]_-$&
$\pi[651\frac{3}{2}]_+$&
$\pi[651\frac{3}{2}]_-$&
$\nu[541\frac{1}{2}]_-$&
$\nu[523\frac{7}{2}]_+$&
$\nu[651\frac{1}{2}]_+$&
$\nu[651\frac{1}{2}]_-$&
$\nu[642\frac{5}{2}]_+$&
$\nu[402\frac{5}{2}]_+$&
$\nu[402\frac{5}{2}]_-$   \\ [2pt]
\tableline
\vspace{-6pt}
      &    &    &       &     &      &     &   \\[1pt]
$^{147}$Eu(1) &$0.153\hspace{9pt}$& &$0.817\hspace{9pt}$&$-0.204\hspace{9pt}$& & &$0.731\hspace{9pt}$&$-0.022\hspace{9pt}$&
                      $-1.394\hspace{9pt}$&$0.609\hspace{9pt}$&$-0.583\hspace{9pt}$ \\ [1pt]
$^{147}$Eu(3) &            & &$0.824\hspace{9pt}$&$-0.341\hspace{9pt}$& & &$0.837\hspace{9pt}$&$0.041\hspace{9pt}$&
                       $-1.494\hspace{9pt}$&$0.623\hspace{9pt}$&$-0.597\hspace{9pt}$ \\ [1pt]
$^{148}$Eu(1) &$0.166\hspace{9pt}$& &$0.834\hspace{9pt}$&$-0.353\hspace{9pt}$& & &$0.856\hspace{9pt}$&$0.028\hspace{9pt}$&
                       $-1.521\hspace{9pt}$&$0.614\hspace{9pt}$&$-0.588\hspace{9pt}$  \\ [1pt]
$^{147}$Gd(2) &$-0.012\hspace{9pt}$&  &$0.848\hspace{9pt}$&$-0.183\hspace{9pt}$& & &$0.765\hspace{9pt}$&$-0.070\hspace{9pt}$&
                       $-1.442\hspace{9pt}$&$0.588\hspace{9pt}$&$-0.561\hspace{9pt}$  \\ [1pt]
$^{147}$Gd(3) &$0.081\hspace{9pt}$& &$0.767\hspace{9pt}$&$0.175\tablenote{Strongly disturbed
  orbital}\hspace{4.7pt}$&$0.143\hspace{9pt}$&$-0.402\hspace{9pt}$& &$-0.159\hspace{9pt}$&$-1.180\hspace{9pt}$&$0.597\hspace{9pt}$&$-0.570\hspace{9pt}$  \\ [1pt]
$^{147}$Gd(4) &$0.064\hspace{9pt}$& &$0.859\hspace{9pt}$&$-0.321\hspace{9pt}$& & &$0.872\hspace{9pt}$&$-0.007\hspace{9pt}$&
                       $-1.549\hspace{9pt}$&$0.560\hspace{9pt}$&$-0.573\hspace{9pt}$  \\ [1pt]
$^{148}$Gd(1) &$0.044\hspace{9pt}$& &$0.842\hspace{9pt}$&$-0.231\hspace{9pt}$& & &$0.790\hspace{9pt}$&$-0.042\hspace{9pt}$&
                       $-1.463\hspace{9pt}$&$0.597\hspace{9pt}$&$-0.570\hspace{9pt}$  \\ [1pt]
$^{148}$Gd(3) &$0.091\hspace{9pt}$& &$0.801\hspace{9pt}$&$-0.038\hspace{9pt}$& & & &$-0.090\hspace{9pt}$&
                       $-1.303\hspace{9pt}$&$0.602\hspace{9pt}$&$-0.575\hspace{9pt}$ \\[1pt]
$^{148}$Gd(4) &$0.006\hspace{9pt}$& &$0.865\hspace{9pt}$&$-0.334\hspace{9pt}$& & &$0.888\hspace{9pt}$&$-0.015\hspace{9pt}$&
                       $-1.570\hspace{9pt}$&$0.594\hspace{9pt}$&$-0.567\hspace{9pt}$ \\[1pt]
$^{148}$Gd(5) & & &$0.877\hspace{9pt}$&$-0.673\hspace{9pt}$& & &$-0.664\hspace{9pt}$& &
                          $0.071\hspace{9pt}$&$0.639\hspace{9pt}$&$-0.613\hspace{9pt}$ \\[1pt]
$^{148}$Gd(6) &$0.080\hspace{9pt}$&$0.290\hspace{9pt}$&$0.873\hspace{9pt}$&$-0.449\hspace{9pt}$& & &$0.976\hspace{9pt}$&$0.042\hspace{9pt}$&
                       $-1.652\hspace{9pt}$&$0.605\hspace{9pt}$&$-0.578\hspace{9pt}$ \\[1pt]
$^{149}$Gd(1) &$0.061\hspace{9pt}$& &$0.858\hspace{9pt}$&$-0.376\hspace{9pt}$& & &$0.914\hspace{9pt}$&$0.011\hspace{9pt}$&
                       $-1.591\hspace{9pt}$&$0.603\hspace{9pt}$&$-0.576\hspace{9pt}$ \\[1pt]
$^{150}$Gd(1) &$0.058\hspace{9pt}$&$0.726\hspace{9pt}$&$0.886\hspace{9pt}$&$-0.558\hspace{9pt}$& & &$0.947\hspace{9pt}$&$0.084\hspace{9pt}$&
                       $-1.620\hspace{9pt}$&$0.606\hspace{9pt}$&$-0.579\hspace{9pt}$ \\[1pt]
$^{150}$Gd(2) & & &$0.862\hspace{9pt}$&$-0.637\hspace{9pt}$& & &$0.372\hspace{9pt}$&$0.167\hspace{9pt}$&
                       $-0.991\hspace{9pt}$&$0.640\hspace{9pt}$&$-0.613\hspace{9pt}$ \\[1pt]
$^{150}$Gd(4b) &$-0.157\hspace{9pt}$& &$0.892\hspace{9pt}$&$-0.422\hspace{9pt}$& & &$0.986\hspace{9pt}$&$-0.012\hspace{9pt}$&
                       $-1.683\hspace{9pt}$&$0.578\hspace{9pt}$&$-0.551\hspace{9pt}$ \\[1pt]
$^{150}$Gd(8a) &$0.008\hspace{9pt}$&$0.401\hspace{9pt}$&$0.881\hspace{9pt}$&$-0.491\hspace{9pt}$& & &$1.004\hspace{9pt}$&$0.045\hspace{9pt}$&
                       $-1.685\hspace{9pt}$&$0.599\hspace{9pt}$&$-0.572\hspace{9pt}$ \\[1pt]
$^{151}$Gd(1a,1b) &$-0.165\hspace{9pt}$&$0.769\hspace{9pt}$&$0.919\hspace{9pt}$&$-0.587\hspace{9pt}$& & &$0.989\hspace{9pt}$&$0.065\hspace{9pt}$&
                       $-1.684\hspace{9pt}$&$0.582\hspace{9pt}$&$-0.554\hspace{9pt}$ \\[1pt]
$^{151}$Tb(1) &$0.089\hspace{9pt}$& &$0.892\hspace{9pt}$&$-0.606\hspace{9pt}$& & &$0.764\hspace{9pt}$&$0.113\hspace{9pt}$&
                       $-1.427\hspace{9pt}$&$0.615\hspace{9pt}$&$-0.588\hspace{9pt}$ \\ [1pt]
$^{152}$Dy(1) &$0.077\hspace{9pt}$&$-0.356\hspace{9pt}$&$0.908\hspace{9pt}$&$-0.666\hspace{9pt}$& & &$0.408\hspace{9pt}$&$0.149\hspace{9pt}$&
                       $-1.057\hspace{9pt}$&$0.620\hspace{9pt}$&$-0.592\hspace{9pt}$ \\ [1pt]
\end{tabular}
\label{tab4}
\end{table*}
This finding explains the
observation of the staggering effect in identical SD bands
\cite{Has2}. An exception is the band $^{148}$Gd(5),
which exhibits clear evidence for staggering. Its configuration
is the same as those for the bands $^{150}$Gd(2) or
$^{152}$Dy(1) apart from two neutron holes in the state
$[411]1/2$ or the two $[411]1/2$ neutron and two
$[301]1/2$ proton holes correspondingly.
Nevertheless, statistically significant staggering has not been
observed in the latter bands. One would suppose that the
superposition principle does not work in this case. This suggestion
is confirmed by the large nonaxial deformation of $^{148}$Gd(5)
found in the calculations of Ref.\ \cite{Karl1}.

The active orbitals give us a more rigorous verification of the
theory. A nucleon occupying this state adds significantly to the
quantity $Q_{44}$ and may change its sign. Table\ \ref{tab4}
shows some active orbitals involving in the configurations of
almost all the studied bands and the estimated values $q_{44}$
for them. The orbital $\nu[651]1/2\ (\alpha=1/2)$  is one such
example. Starting with the staggering bands $^{148}$Eu(1),
$^{148}$Gd(6), $^{149}$Gd(1) and removing a neutron from
this orbital we get correspondingly the bands $^{147}$Eu(1),
$^{147}$Gd(4), $^{148}$Gd(1), which do not stagger. Thus
this active orbital explains the remarkable property of the
$\Delta I=4$ bifurcation observed in Ref.\ \cite{Has}.

In the next step we consider the signature partner bands based
on the state $\nu[402]5/2$, which are associated with the
generation of identical bands. This active orbital has the
reasonably large values of the moment $q_{44}$ to modify the
inequality (\ref{cond}). Consequently a pair of identical bands
may have different staggering properties. The example is the
band $^{150}$Gd(4b), which is identical to $^{149}$Gd(1) but
does not exhibit staggering because the state
$\nu[402]5/2\ (\alpha=-1/2)$ has the large negative value
$q_{44}$. Its signature partner, $^{150}$Gd(4b),
should stagger. Other examples of the signature partner bands
involving this state are shown in Table\ \ref{tab1} and \ref{tab2}.

We extend now this procedure to the bands involving the
configurations that differ by an arbitrary number of particles and
holes in active and inactive orbitals. For the fixed rotational
frequency, the $Q_{44}$ values of the two bands A and B are
connected by the equality
\begin{equation}
Q_{44}({\rm A})=Q_{44}({\rm B})+  \delta Q_{44} +
      \delta Q^{\rm def}_{44},
\label{equa}
\end{equation}
where $\delta Q_{44}$ is the contribution resulting from difference
in active orbitals, while $\delta Q^{\rm def}_{44}$
represents the contribution due to the deformation change, which
induced both active and inactive orbitals. According to additivity
of multipole moments, the former quantity can be written as
\begin{equation}
\delta Q_{44}=\sum_\lambda q_{44}(\lambda),
\label{dqu}
\end{equation}
where $\lambda$ runs over the active particle and/or hole states,
which define the intrinsic configuration of the band A with respect to the
band B (the reference band). 
\begin{table}[t]
\caption{Test of the staggering criterion for the fixed rotational
frequency $\omega$=0.8\, MeV by employing the relative
moment $\Delta Q_{44}$  of the band A with respect to
the reference band B. The symbol $+$ or $-$ is
used to show if the staggering significance of the band A
agrees or disagrees with the sign of $Q_{44}({\rm A})$
obtained from Eqs.\ \ (\ref{equa}) and (\ref{ineqs}). A
blank space means that such comparison is impossible.
It is assumed that the sign of $Q_{44}({\rm B})$ is unchanged
if $|\Delta Q_{44}|<0.5$.  }

\begin{tabular}{|c|cccc|ccccc|ccc|} 
\multicolumn{1}{|l|}{
\begin{picture}(38,40)(0,0) 
\put(6.0,5.0){\Large{A}}
\put(26.0,23.0){\Large{B}}
\put(-1.7,41.0){\line(1,-1){44.0}} 
\put(-1.7,40.7){\line(1,-1){44.0}} 
\end{picture}}
& \rotatebox{90}{$^{148}$Eu(1)}
& \rotatebox{90}{$^{148}$Gd(6)}
& \rotatebox{90}{$^{149}$Gd(1)}
& \rotatebox{90}{$^{151}$Gd(1a)}

& \rotatebox{90}{$^{147}$Gd(2)}
& \rotatebox{90}{$^{147}$Gd(3)}
& \rotatebox{90}{$^{148}$Gd(3)}
& \rotatebox{90}{$^{150}$Gd(2)}
& \rotatebox{90}{$^{150}$Gd(8a)}

& \rotatebox{90}{$^{147}$Gd(4)}
& \rotatebox{90}{$^{148}$Gd(1)}
& \rotatebox{90}{$^{151}$Gd(1b)}
\\ \tableline

\raisebox{0.3 \height}{$^{148}$Eu(1)}  &$\hspace{-0.9mm}$\rule[-1.2mm]{6mm}{6mm}$\hspace{-0.7mm}$ &\raisebox{0.5 \height}{\bf{+}}&\raisebox{0.5 \height}{\bf{+}}& &\raisebox{0.5 \height}{\bf{$-$}}&\raisebox{0.5 \height}{\bf{$-$}}&\raisebox{0.5 \height}{\bf{$-$}}&\raisebox{0.5 \height}{\bf{$-$}}&\raisebox{0.5 \height}{\bf{$-$}}& & & \\  
\raisebox{0.3 \height}{$^{148}$Gd(6)}  &\raisebox{0.5 \height}{\bf{+}}&$\hspace{-0.9mm}$\rule[-1.2mm]{6mm}{6mm}$\hspace{-0.7mm}$ &\raisebox{0.5 \height}{\bf{+}}&\raisebox{0.5 \height}{\bf{+}}&\raisebox{0.5 \height}{\bf{$-$}}&\raisebox{0.5 \height}{\bf{$-$}}&\raisebox{0.5 \height}{\bf{$-$}}&\raisebox{0.5 \height}{\bf{$-$}}&\raisebox{0.5 \height}{\bf{$-$}}& & &  \\ 
\raisebox{0.3 \height}{$^{149}$Gd(1)}  &\raisebox{0.5 \height}{\bf{+}}&\raisebox{0.5 \height}{\bf{+}}&$\hspace{-0.9mm}$\rule[-1.2mm]{6mm}{6mm}$\hspace{-0.7mm}$ & &\raisebox{0.5 \height}{\bf{$-$}}&\raisebox{0.5 \height}{\bf{$-$}}&\raisebox{0.5 \height}{\bf{$-$}}&\raisebox{0.5 \height}{\bf{$-$}}&\raisebox{0.5 \height}{\bf{$-$}}& & &  \\ 
\raisebox{0.3 \height}{$^{151}$Gd(1a)} &\raisebox{0.5 \height}{\bf{+}}&\raisebox{0.5 \height}{\bf{+}}&\raisebox{0.5 \height}{\bf{+}}&$\hspace{-0.9mm}$\rule[-1.2mm]{6mm}{6mm}$\hspace{-0.7mm}$ &\raisebox{0.5 \height}{\bf{$-$}}& & &\raisebox{0.5 \height}{\bf{$-$}}& & & &  \\ \tableline

\raisebox{0.3 \height}{$^{147}$Gd(2)}  &\raisebox{0.5 \height}{\bf{$-$}}&\raisebox{0.5 \height}{\bf{$-$}}&\raisebox{0.5 \height}{\bf{$-$}}&\raisebox{0.5 \height}{\bf{$-$}}&$\hspace{-0.9mm}$\rule[-1.2mm]{6mm}{6mm}$\hspace{-0.7mm}$ & & &\raisebox{0.5 \height}{\bf{+}}& & & &  \\ 
\raisebox{0.3 \height}{$^{147}$Gd(3)}  &\raisebox{0.5 \height}{\bf{$-$}}&\raisebox{0.5 \height}{\bf{$-$}}&\raisebox{0.5 \height}{\bf{$-$}}& &\raisebox{0.5 \height}{\bf{+}}&$\hspace{-0.9mm}$\rule[-1.2mm]{6mm}{6mm}$\hspace{-0.7mm}$ &\raisebox{0.5 \height}{\bf{+}}&\raisebox{0.5 \height}{\bf{+}}&\raisebox{0.5 \height}{\bf{+}}& & &\raisebox{0.5 \height}{\bf{+}} \\ 
\raisebox{0.3 \height}{$^{148}$Gd(3)}  &\raisebox{0.5 \height}{\bf{$-$}}&\raisebox{0.5 \height}{\bf{$-$}}&\raisebox{0.5 \height}{\bf{$-$}}& &\raisebox{0.5 \height}{\bf{+}}&\raisebox{0.5 \height}{\bf{+}}&$\hspace{-0.9mm}$\rule[-1.2mm]{6mm}{6mm}$\hspace{-0.7mm}$ &\raisebox{0.5 \height}{\bf{+}}&\raisebox{0.5 \height}{\bf{+}}& & &\raisebox{0.5 \height}{\bf{+}} \\ 
\raisebox{0.3 \height}{$^{150}$Gd(2)}  &\raisebox{0.5 \height}{\bf{$-$}}&\raisebox{0.5 \height}{\bf{$-$}}&\raisebox{0.5 \height}{\bf{$-$}}&\raisebox{0.5 \height}{\bf{$-$}}&\raisebox{0.5 \height}{\bf{+}}&\raisebox{0.5 \height}{\bf{+}}&\raisebox{0.5 \height}{\bf{+}}&$\hspace{-0.9mm}$\rule[-1.2mm]{6mm}{6mm}$\hspace{-0.7mm}$ &\raisebox{0.5 \height}{\bf{+}}& & &  \\ 
\raisebox{0.3 \height}{$^{150}$Gd(8a)} &\raisebox{0.5 \height}{\bf{$-$}}&\raisebox{0.5 \height}{\bf{$-$}}&\raisebox{0.5 \height}{\bf{$-$}}& &\raisebox{0.5 \height}{\bf{+}}&\raisebox{0.5 \height}{\bf{+}}&\raisebox{0.5 \height}{\bf{+}}&\raisebox{0.5 \height}{\bf{+}}&$\hspace{-0.9mm}$\rule[-1.2mm]{6mm}{6mm}$\hspace{-0.7mm}$ & & &  \\ \tableline

\raisebox{0.3 \height}{$^{147}$Gd(4)}  & & & & &\raisebox{0.5 \height}{\bf{+}}&\raisebox{0.5 \height}{\bf{+}}&\raisebox{0.5 \height}{\bf{+}}&\raisebox{0.5 \height}{\bf{+}}&\raisebox{0.5 \height}{\bf{+}}&$\hspace{-0.9mm}$\rule[-1.2mm]{6mm}{6mm}$\hspace{-0.7mm}$ &\raisebox{0.5 \height}{\bf{+}}&\raisebox{0.5 \height}{\bf{+}} \\ 
\raisebox{0.3 \height}{$^{148}$Gd(1)}  & & & & &\raisebox{0.5 \height}{\bf{+}}&\raisebox{0.5 \height}{\bf{+}}&\raisebox{0.5 \height}{\bf{+}}&\raisebox{0.5 \height}{\bf{+}}&\raisebox{0.5 \height}{\bf{+}}&\raisebox{0.5 \height}{\bf{+}}&$\hspace{-0.9mm}$\rule[-1.2mm]{6mm}{6mm}$\hspace{-0.7mm}$ &\raisebox{0.5 \height}{\bf{+}} \\ 
\raisebox{0.3 \height}{$^{151}$Gd(1b)} & & & & &\raisebox{0.5 \height}{\bf{+}}&\raisebox{0.5 \height}{\bf{+}}&\raisebox{0.5 \height}{\bf{+}}&\raisebox{0.5 \height}{\bf{+}}&\raisebox{0.5 \height}{\bf{+}}&\raisebox{0.5 \height}{\bf{+}}&\raisebox{0.5 \height}{\bf{+}}&$\hspace{-0.9mm}$\rule[-1.2mm]{6mm}{6mm}$\hspace{-0.7mm}$   \\ 
\end{tabular}
\label{tab5}
\end{table}
Since the contributions
$\delta Q_{44}$ and $\delta Q^{\rm def}_{44}$
may be comparable, we have used the values
$Q_{44}$ listed in Tables \ref{tab1} and \ref{tab2} to evaluate the
relative nonaxial moment of active orbitals
\begin{equation}
\Delta Q_{44}= \delta Q_{44} + \delta Q^{\rm def}_{44} =
     Q_{44}({\rm A})-Q_{44}({\rm B}).
\label{deq}
\end{equation}
These quantities along with the staggering significances $Y_{\rm A}$
and $Y_{\rm B}$ allow us to get the more sophisticated test of the
staggering criterion.

We have first selected twelve bands having the proper staggering
significances to deal with the sample involving staggering
($Y\!\geq\!1.8$) and nonstaggering ($Y\!\leq\!0.25$)
bands with a reasonable high likelihood. According to Eqs.
(\ref{cond}) and (\ref{qu}), the former are characterized by the
value $Q_{44}>0$ and the latter have $Q_{44}<0$. To compare
the staggering properties of the bands A and B, we consider
the two strong inequalities
$$
\hspace{-18pt}  Q_{44}({\rm B}) + \Delta Q_{44}>0,\hspace{2mm}
              {\rm if}\hspace{2mm} Y_{\rm B}\geq 1.8,\ \Delta Q_{44}>0,
$$
\begin{equation}
Q_{44}({\rm B}) + \Delta Q_{44}<0,\hspace{2mm}
        {\rm if}\hspace{2mm} Y_{\rm B}\leq 0.25,\ \Delta Q_{44}<0.
\label{ineqs}
\end{equation}
The transformation $\Delta Q_{44}\rightarrow -\Delta Q_{44}$
makes the sign of the sum $Q_{44}({\rm B}) + \Delta Q_{44}$ in
the inequalities (\ref{ineqs}) indefinite unless the absolute value
of $\Delta Q_{44}$ is small compared to $Q_{44}({\rm B})$.
Let us consider, for example, the bands $^{148}$Gd(1)
with $Y_{\rm A}=0.23$ and $^{147}$Gd(3) with $Y_{\rm B}=0.25$,
for which $\Delta Q_{44}=-0.71$. According to Eq.\ (\ref{equa}) and
the second inequality (\ref{ineqs}), we have $Q_{44}({\rm A})<0$
that is in agreement with absence of staggering in the band
$^{148}$Gd(1). On the other hand, considering $^{148}$Gd(1) as
the reference band, we cannot find the staggering behavior
of the band $^{147}$Gd(3) because the sign of the right hand
side of Eq. (\ref{equa}) is indefinite.

The result of such comparison for 132 pairs of bands is presented
in Table \ref{tab5}. The columns of this table involve the reference
bands B, whereas lines represent the bands A. The symbol $+$
($-$) means that Eq.\ (\ref{equa}) and the inequalities (\ref{ineqs})
determine the staggering behavior of the band A correctly
(incorrectly). A blank space is used when the sign of
$Q_{44}({\rm A})$ is indefinite and its comparison with the
significance $Y_{\rm A}$ is impossible. The three  groups of bands
are clearly visible in Table \ref{tab5}. ({\it i}) The nonstaggering
bands $^{147}$Gd(4), $^{148}$Gd(1), $^{151}$Gd(1b). There is
no contradiction in the staggering behavior inside this group of
bands. Such contradiction has not been found also between these
bands and the bands of other groups. ({\it ii}) The four bands with
clear evidence of staggering $^{148}$Eu(1), $^{148}$Gd(6),
$^{149}$Gd(1), and $^{151}$Gd(1a). Whether or not the staggering
behavior of the last band contradicts with that of the band
$^{148}$Eu(1) or $^{149}$Gd(1) is not clear. ({\it iii}) The most striking feature of
Table \ref{tab5} is the third group of bands, which contradict with
all the staggering bands of the second group. 
The bands $^{147}$Gd(3), $^{148}$Gd(3)
with the empty first intruder $\nu 7_1$ and the band $^{150}$Gd(2)
with the blocked first intruder belong to this group. It should be noted that
the band $^{148}$Gd(5), being included in the sample, contradicts with 
the first and third groups.

We have presented a systematic study of the $\Delta I=4$
bifurcation in the SD bands of the $A\sim 150$ mass region. The
analysis is based on the necessary condition of the staggering
phenomenon obtained in previous works and
the MO potential in the pure single-particle approach. The
results show that the criterion is working surprisingly well and is
in a reasonably agreement with the statistical analysis of Haslip
et al. We also have revealed the set of the single-particle states
with a large nonaxial hexadecapole moment (active orbitals). They
clearly highlight the role of the hexadecapole degree of freedom
in the phenomenon and allows to answer the main question of why
the staggering is not a universal feature of SD bands. Another
class of states (inactive orbitals) allows to explain the
observation of staggering in the identical bands $^{148}$Eu(1),
$^{148}$Gd(6), and $^{149}$Gd(1), which are the only bands
clearly exhibiting the effect. Some discrepancies and
contradictions established by the analysis indicate the need
for further experimental and theoretical study 
of this interesting phenomenon.

The authors wish to thank Anatoli Afanasjev, Ingemar Ragnarsson
for supplying the information concerning the structure and
deformations of some bands, and Duncan Appelbe for giving the
$^{150,151}$Gd band configurations.


\begin{references}
\bibitem[*]{IMP}E-mail address: pavi@jbivn.kiae.su
\bibitem{Lal}G. A. Lalazissis and K. Hara, Phys. Rev. C
{\bf 58}, 243 (1998).
\bibitem{Has}D. S. Haslip, S. Flibotte, C. E. Svensson, and
J. C. Waddington, Phys. Rev. C {\bf 58}, R1893 (1998).
\bibitem{Has1} D. S. Haslip, {\it et al.}, Phys. Rev. C
{\bf 58}, R2649 (1998).
\bibitem{HM}I. Hamamoto and B. Mottelson, Phys. Lett. {\bf B333},
294 (1994); Phys. Scripta {\bf T56}, 27 (1995).
\bibitem{Mac}A. O. Macchiavelli {\it et al.}, Phys. Rev. C
{\bf 51}, R1 (1995).
\bibitem{PF}I. M. Pavlichenkov and S. Flibotte, Phys. Rev. C
{\bf 51}, R460 (1995).
\bibitem{Mag}P. Magierski, K. Burzy\'nski and J. Dobaczewski,
Acta Phys. Polonica B {\bf 26} (1995) 291.
\bibitem{MQ}I. N. Mikhailov and P. Quentin, Phys. Rev. Lett.
{\bf 74}, 3336 (1995).
\bibitem{Yan}Yang Sun, Jing-ye Zhang and M. Guidry, Phys. Rev. Lett.
{\bf 75}, 3398 (1995).
\bibitem{Pav}I. M. Pavlichenkov, Phys. Rev. C,
{\bf 55}, 1275 (1996).
\bibitem{Pav1}I. M. Pavlichenkov, JETP Lett.
{\bf 66}, 796 (1997).
\bibitem{Ragnar} I. Ragnarsson and P. B. Semmes, Hyperfine
Interactions, {\bf 43}, 425 (1988).
\bibitem{Haas}B. Haas {\it et al.}., Nucl. Phys. {\bf A561}, 251 (1993).
\bibitem{Hebb}G. Hebbinghaus {\it et al.}., Phys. Lett. {\bf B240},
311 (1990).
\bibitem{Sat}W. Satula, J. Dobaczewski, J. Dudek and W. Nazarewicz,
Phys. Rev. Lett. {\bf 77}, 5182 (1996).
\bibitem{Rag}I. Ragnarsson, Poster of the Conference on Physics from
Large $\gamma$-Ray Detector Arrays, Berkeley, 1994.
\bibitem{Naz}P. Magierski, P.-H. Heenen and W. Nazarewicz,
Phys. Rev. C {\bf 51}, R2880 (1995).
\bibitem{Thei}Ch. Theiseh {\it et al.}, Phys. Rev. C
{\bf 54}, 2910 (1996).
\bibitem{Sav}H. Savajols {\it et al.}, Phys. Rev. Lett. {\bf 76}, 4480 (1996).
\bibitem{Rag1}I. Ragnarsson, private communication.
\bibitem{Flib}S. Flibotte {\it et al.}., Nucl. Phys. {\bf A584}, 373 (1995).
\bibitem{Nazar}W. Nazarewicz and I. Ragnarsson, in
{\it Handbook of Nuclear Properties} edited by D. Poenaru and
W. Greiner (Clarendon Press, Oxford 1996), p. 80.
\bibitem{deFra}G. de France {\it et al.}., Phys. Lett. {\bf B331},
290 (1994).
\bibitem{Karl}L. B. Karlsson, I. Ragnarsson, and S. \AA berg,
Nucl. Phys. {\bf A639}, 654 (1998).
\bibitem{Flibo} S. Flibotte, {\it et al.}, Phys. Rev. Lett.
{\bf 71}, 4299 (1993).
\bibitem{Has2} D. S. Haslip, {\it et al.}, Phys. Rev. Lett.
{\bf 78}, 3447 (1997).
\bibitem{Karl1}L. B. Karlsson, Doctoral dissertation, Lund University,
1999.

\end{references}
\end{document}